\documentclass[conference]{IEEEtran}
\PassOptionsToPackage{draft}{hyperref}

\usepackage{xcolor}
\usepackage[linesnumbered,ruled,vlined]{algorithm2e}
\usepackage{bm}
\usepackage{amssymb}
\usepackage{comment}
\usepackage{color}
\usepackage{bm}
\usepackage{tabularx,booktabs}
\newcolumntype{C}{>{\centering\arraybackslash}X} % centered version of "X" type
\usepackage{lipsum}
\usepackage{caption}
\usepackage{subcaption}
\ifCLASSINFOpdf
   \usepackage[pdftex]{graphicx}
  \graphicspath{ {./images/} }
   \DeclareGraphicsExtensions{.pdf,.jpeg,.png}
\else
\fi
\usepackage{amsmath}
\usepackage{mathrsfs}

\usepackage{acro}
\providecommand\tooltip[2]{[PDF:no. of 1][tooltip:no. of 2]}
\acsetup{tooltip=false}

\DeclareAcronym{CS}{
  short        = {CS},
  long         = {current-steering}
}

\DeclareAcronym{FFT}{
  short        = {FFT},
  long         = {Fast Fourier Transform}
}

\DeclareAcronym{DnC}{
  short        = {DnC},
  long         = {Deep-n-Cheap}
}

\DeclareAcronym{ReLu}{
  short        = {ReLu},
  long         = {rectified linear unit}
}

\DeclareAcronym{MSE}{
  short        = {MSE},
  long         = {mean squared error}
}

\DeclareAcronym{DAC}{
  short        = {DAC},
  long         = {digital-to-analog converter}
}

\DeclareAcronym{ADC}{
  short        = {ADC},
  long         = {analog-to-digital converter}
}

\DeclareAcronym{DPD}{
  short        = {DPD},
  long         = {digital pre-distortion}
}

\DeclareAcronym{LUT}{
  short        = {LUT},
  long         = {lookup-table}
}

\DeclareAcronym{TDNN}{
  short        = {TDNN},
  long         = {time-delay neural network}
}

\DeclareAcronym{MLP}{
  short        = {MLP},
  long         = {multi-layer-perceptron}
}

\DeclareAcronym{NN}{
  short        = {NN},
  long         = {neural network}
}

\DeclareAcronym{DEM}{
  short        = {DEM},
  long         = {dynamic element matching}
}

\DeclareAcronym{SGD}{
  short        = {SGD},
  long         = {stochastic gradient descent}
}

\DeclareAcronym{IM}{
  short        = {IM},
  long         = {intermodulation}
}

\DeclareAcronym{FIR}{
  short        = {FIR},
  long         = {finite impulse response}
}

\begin{document}
\title{Linearization for High-Speed Current-Steering DACs Using Neural Networks \\ 
\thanks{This work was supported in part by the National Science Foundation (CCF-1763747, ECCS 1643004) and Jariet Technologies. }}

%\author{\IEEEauthorblockN{Daniel Beauchamp}
%\IEEEauthorblockA{
%Ming Hsieh Department of Electrical Engineering\\
%University of Southern California,\\ Los Angeles, California 90089\\
%Email: dbeaucha@usc.edu}
%\and
%\IEEEauthorblockN{Keith M. Chugg}
%\IEEEauthorblockA{Ming Hsieh Department of Electrical Engineering\\
%University of Southern California,\\ Los Angeles, California 90089\\
%Email: chugg@usc.edu}
%}

\author{
    \IEEEauthorblockN{
        Daniel Beauchamp\IEEEauthorrefmark{1}\IEEEauthorrefmark{2} and
        Keith M. Chugg\IEEEauthorrefmark{2}}
    \IEEEauthorblockA{
        \begin{tabular}{cc}
            \begin{tabular}{@{}c@{}}
                \IEEEauthorrefmark{1}
                    Jariet Technologies, 103 W Torrance Blvd, Redondo Beach, CA 90277 \\
                \IEEEauthorrefmark{2}
                  Ming Hsieh Department of Electrical Engineering, \\
University of Southern California, Los Angeles, California 90089 \\
                    \{dbeaucha, chugg\}@usc.edu
            \end{tabular} 
        \end{tabular}
    }
}


%Telephone: (800) 555--1212, Fax: (888) 555--1212}


\IEEEoverridecommandlockouts
\IEEEpubid{\makebox[\columnwidth]{978-1-7281-7670-3/21/\$31.00~
\copyright2021
IEEE \hfill} \hspace{\columnsep}\makebox[\columnwidth]{ }}

\maketitle

\begin{abstract}
This paper proposes a novel foreground linearization scheme for a high-speed \ac{CS} \ac{DAC}. The technique leverages \acp{NN} to derive a \ac{LUT} that maps the inverse of the \ac{DAC} transfer characteristic onto the input codes. The algorithm is shown to improve conventional methods by at least 6dB in terms of \ac{IM} performance for frequencies up to 9GHz on a state-of-the-art 10-bit \ac{CS}-\ac{DAC} operating at 40.96GS/s (gigasamples-per-second) in 14nm CMOS. 
\end{abstract}


% \fi
\IEEEpeerreviewmaketitle

\section{Introduction}
Data converters are now operating at several GS/s with high resolution in compact deep-submicron processes. This is paving the way for commercial applications such as 5G cellular communication and automotive radar \cite{bib:5G_arrays}, \cite{bib:automotive_radar_arrays}. However, it is well known that data converter performance degrades due to nonlinear distortion  \cite{bib:chammas}, \cite{bib:razavi}, which makes modeling and linearization critical. In this paper, we focus on linearization for a high-speed \ac{CS}-\ac{DAC}.

\begin{figure}[ht]
\centering
\includegraphics[scale=0.6]{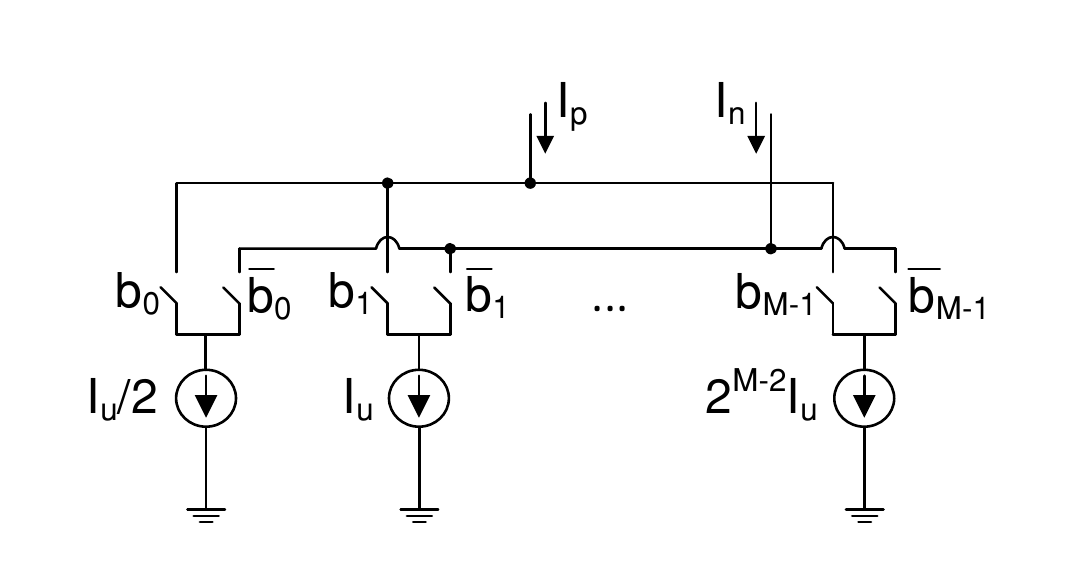}
\caption{Circuit diagram of the $M$-bit \ac{CS} \ac{DAC} with output current $I_\text{out}:=I_p - I_n$.}
\label{fig:high_level_cs_dac}
\end{figure}

Although there are several \ac{DAC} architectures available, the \ac{CS}-\ac{DAC} is regarded as the ``de-facto solution" at gigahertz frequencies \cite{bib:razavi}. A block diagram for the $M$-bit \ac{CS}-\ac{DAC} is shown in Figure \ref{fig:high_level_cs_dac}. It is modeled as an array of binary-weighted current drivers with complementary switching. In reality, the current sources shown in Figure \ref{fig:high_level_cs_dac} differ from their ideal binary weights, and mismatch between them causes large discontinuities in the transfer characteristic thus degrading linearity \cite{bib:razavi}.

In general, the \ac{CS}-\ac{DAC} has both static and dynamic errors. However, in this paper, we consider modern time-interleaved architectures that suppress dynamic errors by hiding code transitions from the output \cite{bib:olieman}. The work in \cite{bib:beauchamp} provides a machine learning-based procedure to calibrate interleaving effects for such architectures.

The focus of this paper is on static nonlinearity, which is mainly attributable to current source mismatch and nonlinear behavior associated with the current drivers. A common remedy is \ac{DEM} which involves randomization over the current drivers to average out mismatch, but this also raises the noise floor. An alternative that does not raise the noise floor is \ac{DPD}. This technique cancels out the nonlinearity by mapping the inverse of the transfer characteristic onto the input codes.

In this paper, we propose a novel \ac{DPD} scheme that is tailored to the discontinuities of the \ac{CS}-\ac{DAC} transfer characteristic. We begin by exciting the \ac{DAC} with an input waveform, and then capturing its output with an \ac{ADC}. Since our scheme is not intended to update in the background, the \ac{DAC} input signal can be designed. We use the term \textit{background} to refer to a scheme that runs during normal operation using \ac{DAC} input data driven by the application.  This is in contrast to a \textit{foreground} scheme which runs offline calibration and allows one to select the \ac{DAC} input data to be used for system identification. In our approach, we design the \ac{DAC} input signal so that it does not  stimulate the dynamic effects in the \ac{DAC} output driver and measurement path from the \ac{DAC} output to the \ac{ADC} input. Thus, only the static transfer characteristic will be identified using the resulting captured input-output pairs.  Specifically, we excite the \ac{DAC} with a low-frequency sine wave so that the static nonlinearity is extracted directly. The static transfer characteristic is then learned by training a \ac{NN} using a dataset of input-output pairs from this \ac{DAC}-to-\ac{ADC} system. Lastly, the inverse of this transfer characteristic is then mapped onto the input codes using a \ac{LUT}, thus linearizing the \ac{DAC}.

% meaning  Our approach is to extract the static nonlinearity of ther \ac{DAC} by exciting  based on \acp{NN}. The justification for using \acp{NN} comes from the fact that the 

% In this paper, we explore \ac{DPD} based on \acp{NN}. Our approach is to to extract the \ac{DAC} static nonlinearity and then first perform system identification by training a \ac{NN}, where the dataset is comprised of \ac{DAC} input codes and corresponding output samples provided by an \ac{ADC}. The inverse of this \ac{DAC}-to-\ac{ADC} system model is then mapped onto the input codes using a \ac{LUT}.  

% Correcting static errors is complicated by the fact that memory is involved in the system, and this is mainly due to finite bandwidth of the \ac{DAC} output driver and \ac{DAC}-to-\ac{ADC} measurement path. However, we found that static errors and memory effects can be isolated from each other by careful design of the datasets, and this promotes memoryless system identification using simple \acp{MLP}. In addition, we use recent tools that provide guidance on how to choose \ac{NN} architectures and hyperparameters as opposed to choosing them heuristically. 

The technique is described in Section II and then simulated in Section III. In Section IV, it is experimentally verified using a state-of-the-art, commercially developed \ac{DAC} operating at 40.96GS/s in 14nm CMOS, to be deployed in end markets such as 5G wireless and advanced radar. Our technique shows an improvement of at least 6dB in terms of \ac{IM} performance compared to conventional \ac{DEM} and polynomial-based \ac{DPD} for frequencies up to 9GHz. We conclude the paper in Section V by summarizing the results.

\section{System Identification}

\begin{figure}[ht]
\centering
\includegraphics[scale=0.7]{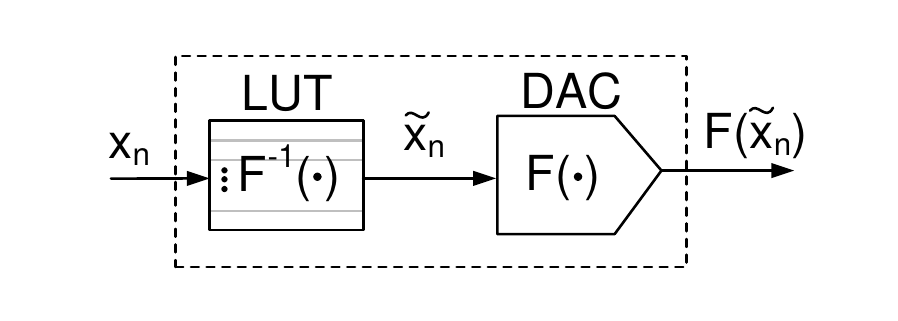}
\caption{Block diagram illustrating the \ac{DPD} concept, where the inverse of the DAC static transfer characteristic is stored in a LUT.}
\label{fig:high_level_dpd}
\end{figure}

\begin{figure}[ht]
\centering
\includegraphics[scale=0.55]{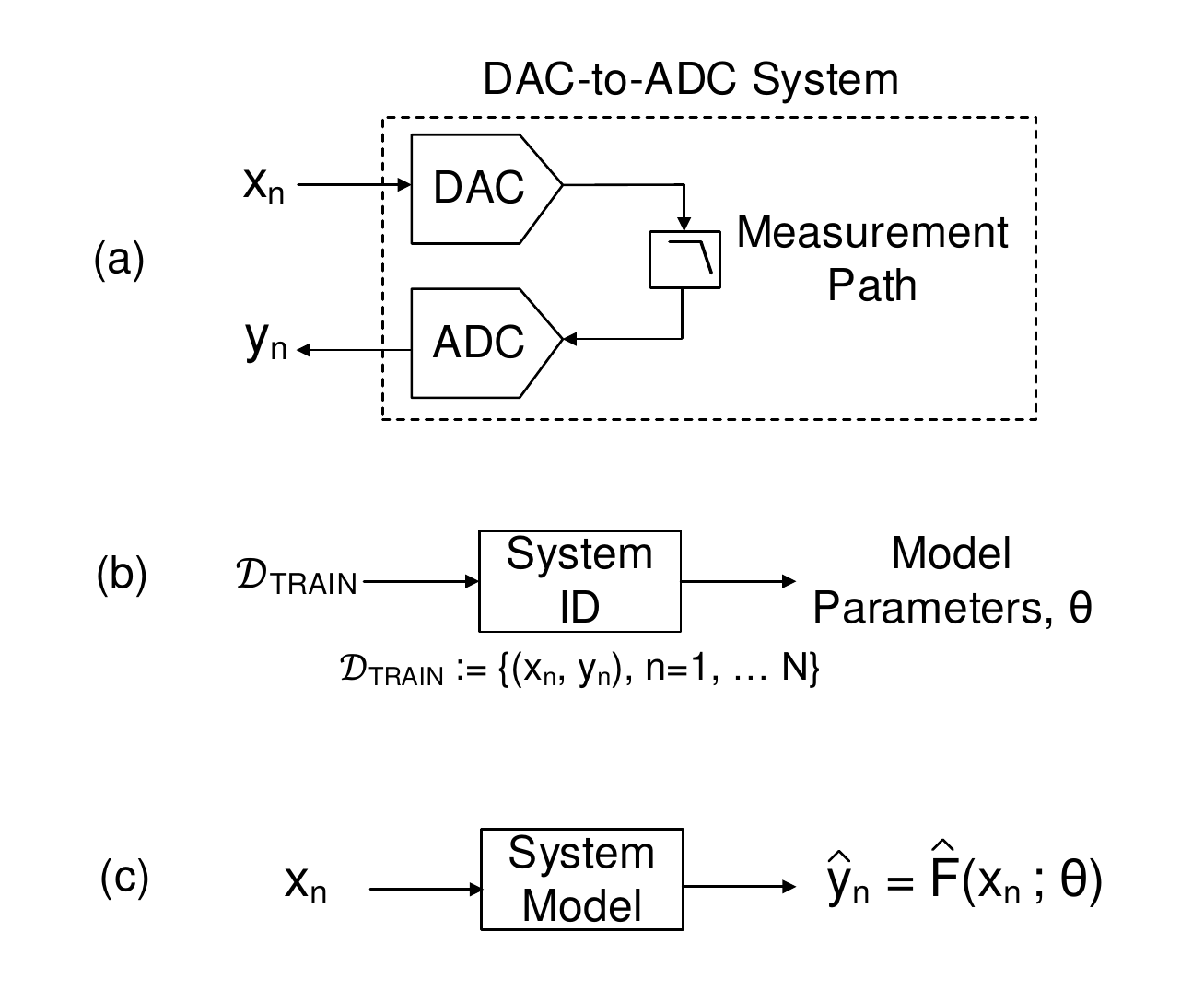}
\caption{(a) Block diagram of the DAC-to-ADC system, (b) System identification using a dataset to determine model parameters $\theta$, (c) DAC-to-ADC system model with input $x_n$ and output $\hat{y}_n$.}
\label{fig:system_id}
\end{figure}

Mapping the \ac{DAC} input codes using \ac{DPD}  in order to remedy static nonlinearity has been investigated in \cite{bib:daigle}, \cite{bib:aleriza}. The main idea is illustrated in Figure \ref{fig:high_level_dpd}, where a \ac{LUT} maps input codes $x_n$ to $\tilde{x}_n = F^{-1}(x_n)$, which linearizes the \ac{DAC} by inverting its static transfer characteristic $F(\cdot)$. The static nonlinearity is modeled as a time-invariant, memoryless system.  We use the term \textit{transfer characteristic} to describe the input-output relationship for this memoryless nonlinearity.
Data from the \ac{DAC} output is required in order to estimate $F(\cdot)$, and this is typically provided by an \ac{ADC}. A block diagram of a representative \ac{DAC}-to-\ac{ADC} system is shown in Figure \ref{fig:system_id}(a), where the measurement path from the \ac{DAC} output to the \ac{ADC} input is modeled as a lowpass filter. Our approach is to obtain an estimate $\hat{F}(\cdot \, ;\theta)$, where $\theta$ are the model parameters. We refer to this as \textit{system identification}, and this is depicted in Figure \ref{fig:system_id}(b) where model parameters $\theta$ are found using a dataset of input-output pairs from the \ac{DAC}-to-\ac{ADC} system: $\mathcal{D}_{\text{TRAIN}} := \left\{(x_n, y_n), \ n=1, \dots, N \right\}$.

The \ac{DAC} stimulus used for system identification in \cite{bib:daigle}, \cite{bib:aleriza} is uniformly distributed random codes. This is done because the proposed algorithms in this case are intended to run in the background, and random codes share spectral properties with the signals encountered during normal operation.  In contrast, we consider a foreground linearization scheme and, consequently, we leverage our choice of input stimulus in order to isolate the static nonlinearity. Specifically, we excite the \ac{DAC} using a sine wave with frequency $f_{\text{sig}} \ll f_s$, where $f_s$ is the \ac{DAC} sample rate. \textit{This avoids stimulating the dynamic effects inherent in the \ac{DAC} output driver and measurement path}. Therefore, we seek a \textit{memoryless} model $\hat{y}_n = \hat{F}(x_n ; \theta)$ as depicted in Figure \ref{fig:system_id}(c). Furthermore, we assume the \ac{ADC} in Figure \ref{fig:system_id}(a) is sufficiently linear so that the \ac{DAC}-to-\ac{ADC} system accurately captures the nonlinearity of the standalone \ac{DAC}.

The choice of the regression model $\hat{F}$ is critical, and depends on the problem at hand. In \cite{bib:daigle} and \cite{bib:aleriza} this model is a polynomial, which is a suitable choice since the proposed \ac{DAC} architecture exhibits only weakly nonlinear behavior. For \ac{CS} architectures, which are the focus of this paper, this model should be selected carefully. This is because the \ac{CS}-\ac{DAC} transfer characteristic is prone to large discontinuities \cite{bib:razavi}. For example, referring to Figure \ref{fig:high_level_cs_dac}, if all current sources are ideal, incrementing the binary input code by 1 produces an output current increase of $I_u$ in all cases.  However, if, for example, the current source corresponding to the most significant bit is $2^{M-2}I_u \, (1+\epsilon)$, the transition from input code $011\cdots 1$ to $100\cdots 0$ will produce a change in output current of $I_u \, (1+\epsilon 2^{M-1})$ instead of the ideal value of $I_u$.  This is the source of jump discontinuities in the transfer characteristic for \ac{CS}-\ac{DAC}s.

\begin{figure}[ht]
\centering
\includegraphics[scale=0.5]{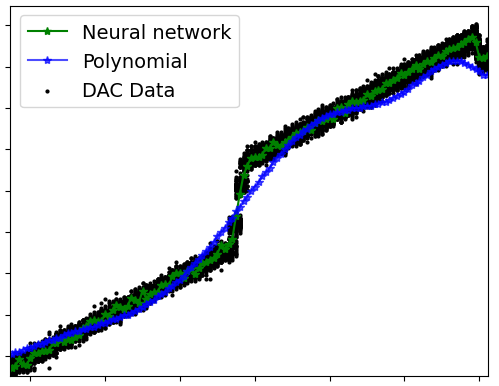}
\caption{Polynomial vs. NN regression in the vicinity of a discontinuity for a CS-DAC behavioral model.}
\label{fig:poly_vs_nn}
\end{figure}

Although polynomials are a popular choice for a regression model, they are ineffective at fitting discontinuities -- i.e., they fit the abrupt transition poorly and exhibit oscillatory behavior \cite{bib:poly_osc_interp}.  In contrast, \ac{NN} regression models are powerful, universal approximators and are a good choice for fitting a transfer characteristic with jump discontinuities as well as other, smooth, nonlinear effects.  This is illustrated in the example shown in Figure \ref{fig:poly_vs_nn} where we have focused on a region of the \ac{CS}-\ac{DAC} transfer characteristic containing a jump discontinuity.  

\begin{figure}[ht]
\centering
\includegraphics[scale=0.7]{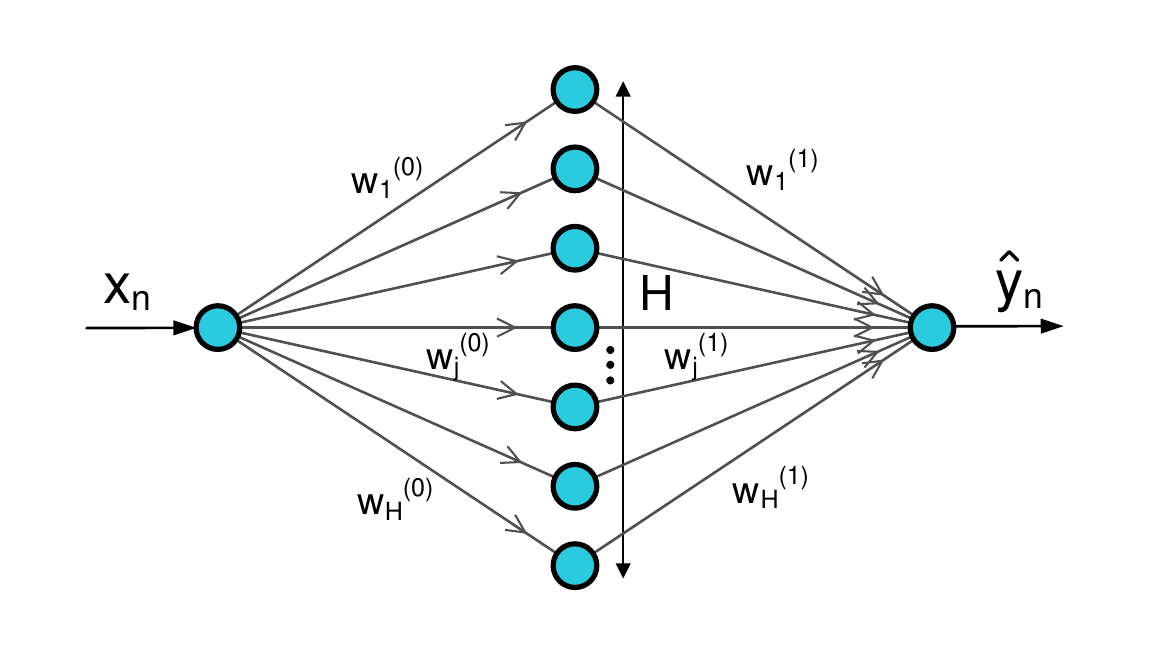}
\caption{Single layer \ac{MLP} with 1 input node, $H$ hidden nodes, and 1 output node.}
\label{fig:mlp_nn}
\end{figure}

Note how the \ac{NN} fits this region well while the polynomial exhibits both poor fitting near the discontinuity and oscillatory behavior. For this reason, we approach system identification using \acp{NN}. The \acp{NN} considered in this paper are feedforward \acp{MLP}. An example of an \ac{MLP} with a single hidden layer is shown in Figure \ref{fig:mlp_nn}, and the output $\hat{y}_n$ for this architecture with nonlinear activation $\underbar{$h$}: \mathbb{R}^{H} \rightarrow \mathbb{R}^{H}$ is given by
\begin{align}
    \hat{y}_n = {\bm{w}^{(1)}}^\top \,  \underbar{$h$} \left( \bm{w}^{(0)} \, x_n + \bm{b}^{(0)} \right)  + b^{(1)}
\end{align}
\noindent where the set of trainable parameters $\theta$ is defined as 
\begin{align}
\theta:= \left\{ \bm{w}^{(0)} , \bm{w}^{(1)}, \bm{b}^{(0)} , b^{(1)}  \right\} 
\end{align}
\noindent with dimensions $\bm{w}^{(0)}\in \mathbb{R}^H$, $\bm{w}^{(1)} \in \mathbb{R}^H$, $\bm{b}^{(0)} \in \mathbb{R}^{H}$, $b^{(1)} \in \mathbb{R}$. 

\section{Simulation Results}
\begin{figure}[ht!]
\includegraphics[scale=0.25]{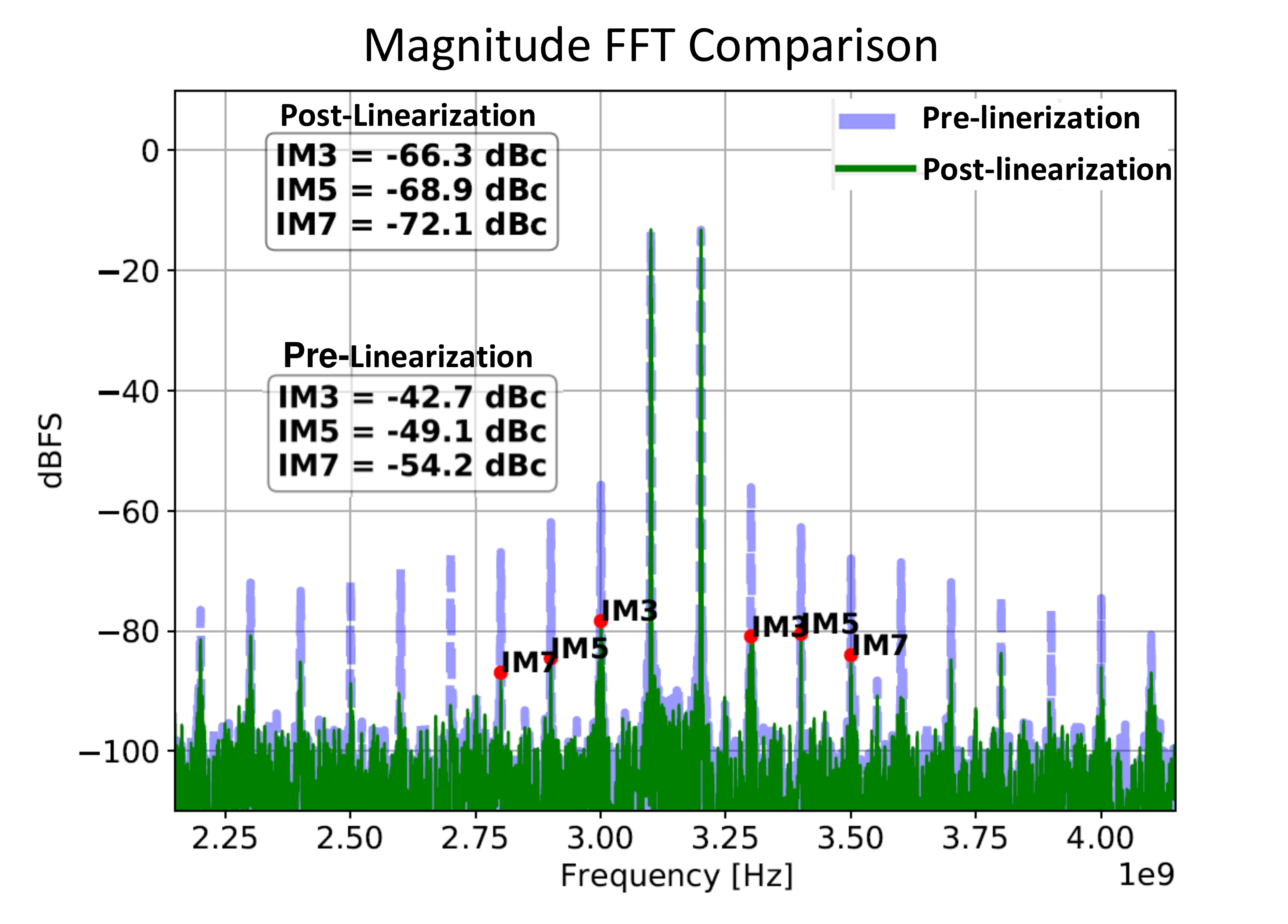}
\caption{Two-tone FFT comparison before and after NN-based DPD. The signal frequencies are $f_1=3.1$GHz, $f_2=3.2$GHz with amplitudes -12dBFS/tone and the DAC is sampling at $f_s=40.96$GS/s. }
\label{fig:im_products}
\end{figure}

In this section, dataset $\mathcal{D}_\text{TRAIN}$ is obtained using 10-bit \ac{DAC} and \ac{ADC} behavioral models operating at $f_s = 40.96 \text{GS/s}$. These MATLAB-based models accurately reflect the behavior of the DAC and ADC used in Section IV. We model the measurement path in Figure \ref{fig:system_id}(a) as a $2^{\text{nd}}$ order Butterworth lowpass filter with 20GHz cutoff. The \ac{FFT} of a two-tone waveform without any linearization is illustrated by the blue spectrum in Figure \ref{fig:im_products}. Note that current source errors result in \ac{IM} products, and the linearization objective is to suppress these as much as possible.

We approach system identification in a \ac{NN} framework by minimizing the following \ac{MSE} cost function

\begin{align}
    C_{\text{model}} = \frac{1}{N} \sum_{n=1}^N \left(\hat{y}_n - y_n \right)^2
\end{align}

\noindent by an appropriate selection of $\theta$, $H$, and $\underbar{$h$} (\cdot)$. Conventionally, hyperparameters $H$, $\underbar{$h$} (\cdot)$, and the number of hidden layers are chosen heuristically. However, in this paper, we leverage \ac{DnC}, an automated framework for low complexity deep learning applications \cite{bib:dnc}. This results in single layer \ac{NN} with \ac{ReLu} activation \cite{bib:relu} and $H = 271$ hidden nodes. Model parameters $\theta$ are then obtained using an extended version of \ac{SGD} \cite{bib:adam}, which completes system identification for the static transfer characteristic. The inverse of this transfer characteristic is then quantized to the 10-bit level and then stored in a \ac{LUT} as shown in Figure \ref{fig:high_level_dpd}.

% \begin{figure}[ht]
% \centering
% \includegraphics[scale=0.35]{images/system_id_fft.pdf}
% \caption{Magnitude FFT comparison of the MLP model and DAC output in dB relative to full scale (dBFS). IM3, IM5, and IM7 products are denoted in dBc (dB-relative-to-carrier).}
% \end{figure}

% In Figure \ref{fig:im_products}, we compare \acp{FFT} of $y_n$ and trained \ac{MLP} predictions $\hat{y}_n$ based on a two-tone input with amplitude -12dBFS/tone and 100 MHz spacing centered at $f_c = 3.15$GHz. Note that current source errors result in \ac{IM} products which are accurately captured by the \ac{MLP} model. Such products serve as a linearity measure, and the calibration objective is consistent with suppressing them as much as possible.

The performance of \ac{NN}-based \ac{DPD} on the behavioral model is illustrated by the green spectrum in Figure \ref{fig:im_products}, which shows a reduction of 23.6dB, 19.8dB, and 17.9dB for IM3, IM5, and IM7 respectively.

\section {Measurement Results}
In this section, we present results for \ac{NN}-based \ac{DPD} on a twofold time-interleaved 10-bit \ac{CS}-\ac{DAC} operating at $f_s = 40.96$GS/s in 14nm CMOS. Our motivation is to demonstrate the ability to capture real-world nonlinearities and also avoid capturing dynamic properties of the system. We do not intend to compare the specific DAC used to state-of-the-art circuit research. 

\begin{figure}[ht]
\centering
\includegraphics[scale=0.035]{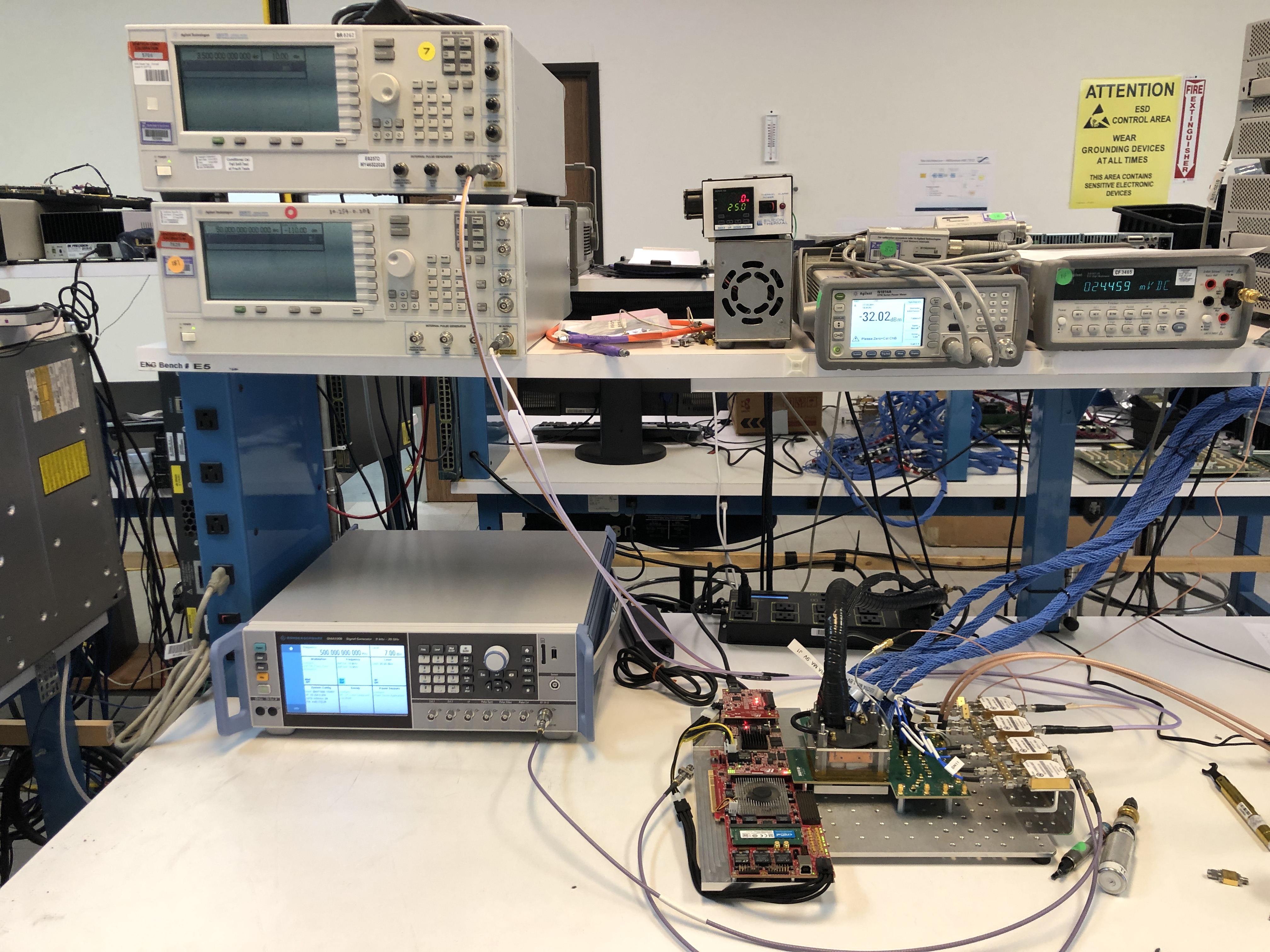}
\caption{Test bench with the high-speed \ac{DAC} and \ac{ADC} test board.}
\label{fig:test_setup}
\end{figure}

Dataset $\mathcal{D}_{\text{TRAIN}}$ is obtained by capturing the \ac{DAC} output using an on-chip 10-bit \ac{ADC} synchronized to the same sample rate as the \ac{DAC}. The \ac{DAC} is externally connected to the \ac{ADC} to avoid undesired signal attenuation and filtering effects. The test setup is shown in Figure \ref{fig:test_setup}. Linearization was performed in the same \ac{NN} framework described in Section III using a sine wave with frequency $f_{\text{sig}}$ = 100 MHz for system identification. 

\begin{figure}[ht]
\centering
\includegraphics[scale=0.25]{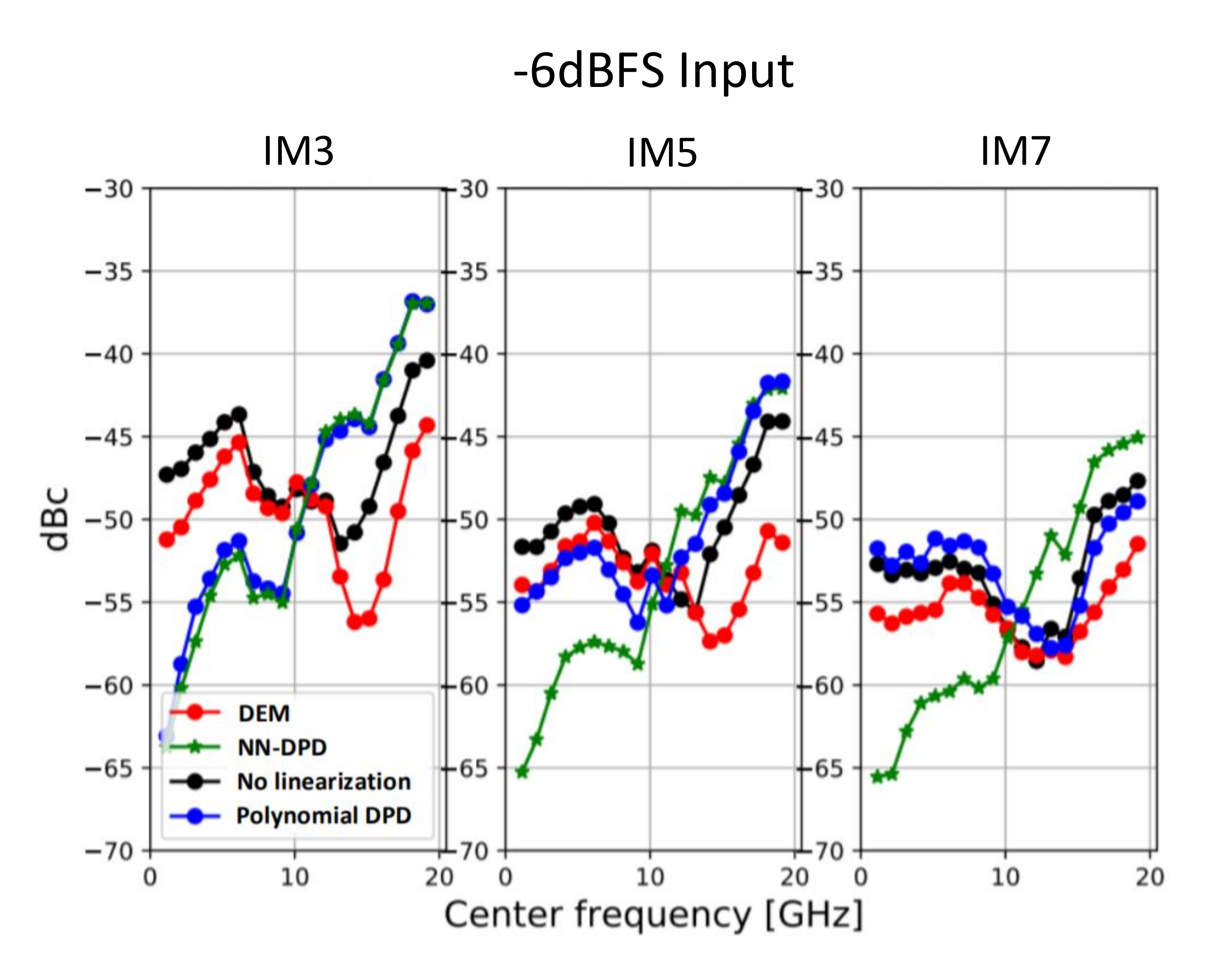}
\caption{IM3/IM5/IM7 performance across Nyquist for two-tone signals, -12dBFS/tone (-6dBFS total amplitude), 100 MHz spacing.}
\label{fig:results_1}
\end{figure}

\begin{figure}[ht]
\centering
\includegraphics[scale=0.22]{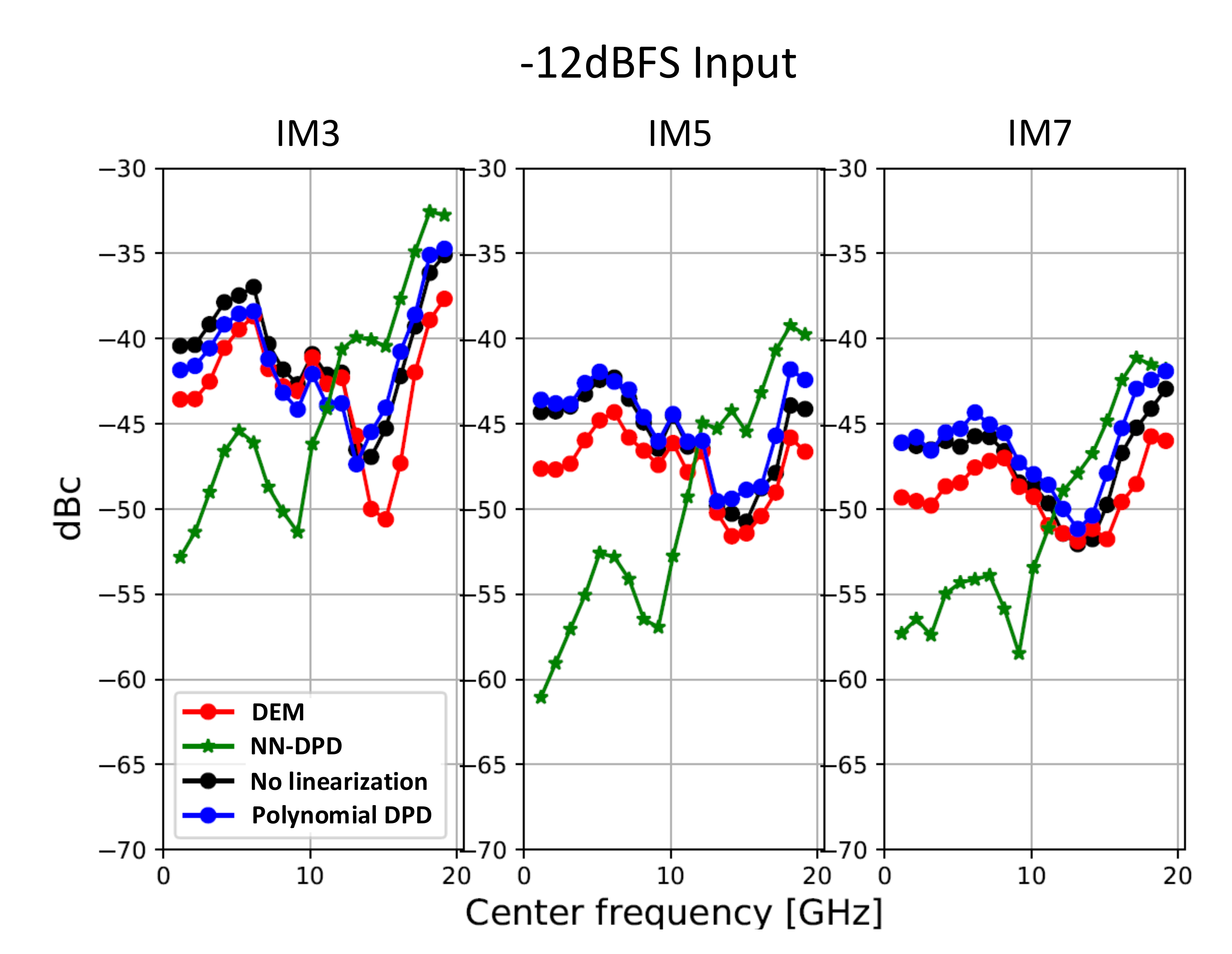}
\caption{IM3/IM5/IM7 performance across Nyquist for two-tone signals, -18dBFS/tone (-12dBFS total amplitude), 100 MHz spacing.}
\label{fig:results_2}
\end{figure}

The results are illustrated in Figure \ref{fig:results_1} and Figure \ref{fig:results_2}, where we compare IM3/IM5/IM7 levels using two-tone signals centered at various frequencies across the first Nyquist zone. System identification is performed with amplitude -6dBFS, and performance is evaluated for both -6dBFS and -12dBFS. We compare the proposed \ac{NN} technique with \ac{DEM} and $15^{\text{th}}$ order polynomial-based \ac{DPD}. An on-chip randomizer is used for the former, and coefficients for the latter are found by applying linear regression with a Vandermonde matrix.

Based on Figure \ref{fig:results_2}, it is evident that \ac{NN}-based \ac{DPD} shows an improvement of at least 6dB for frequencies up to 9GHz for -12dBFS inputs. This is significant for sub-6GHz applications such as 5G. We suspect that pulse shape and timing errors begin to dominate linearity performance above 9GHz. Evidence for this is based on the efficacy of \ac{DEM} above 9GHz, as it is proven to suppress such errors \cite{bib:timing_errors}.

\section {Conclusion}
In this paper, we explored a novel linearization scheme for high-speed current steering \acp{DAC} using \acp{NN}. We showed that simple \acp{MLP} are sufficient for system identification if low-frequency sine waves are used for training. The \ac{NN} architecture is selected using \ac{DnC} and parameters are found using \ac{SGD}. The inverse of the transfer characteristic is then mapped onto the input codes using a \ac{LUT}. The final implementation is a simple pre-distortion \ac{LUT} with no \acp{NN} required. 

A useful extension would be to make this scheme adaptive with respect to temperature and supply voltage variation. This may be accomplished by using sensors coupled with multiple \acp{LUT}. Lastly, our approach demonstrates an improvement of at least 6dB over conventional \ac{DEM} and polynomial-based \ac{DPD} methods for frequencies up to 9GHz.

\section*{Acknowledgment}
We would like to acknowledge Ziping Chen for improving \ac{DnC} by adding the regression feature that was used in this paper.



%\bibliographystyle{IEEEtran}
%\begin{thebibliography}{1}

%H.~Kopka and P.~W. Daly, \emph{A Guide to %\LaTeX}, 3rd~ed.\hskip 1em plus

%\end{thebibliography}
\bibliographystyle{IEEEtran}
\bibliography{bibliography/bibliography}

\begin{thebibliography}{10}
\providecommand{\url}[1]{#1}
\csname url@samestyle\endcsname
\providecommand{\newblock}{\relax}
\providecommand{\bibinfo}[2]{#2}
\providecommand{\BIBentrySTDinterwordspacing}{\spaceskip=0pt\relax}
\providecommand{\BIBentryALTinterwordstretchfactor}{4}
\providecommand{\BIBentryALTinterwordspacing}{\spaceskip=\fontdimen2\font plus
\BIBentryALTinterwordstretchfactor\fontdimen3\font minus
  \fontdimen4\font\relax}
\providecommand{\BIBforeignlanguage}[2]{{%
\expandafter\ifx\csname l@#1\endcsname\relax
\typeout{** WARNING: IEEEtran.bst: No hyphenation pattern has been}%
\typeout{** loaded for the language `#1'. Using the pattern for}%
\typeout{** the default language instead.}%
\else
\language=\csname l@#1\endcsname
\fi
#2}}
\providecommand{\BIBdecl}{\relax}
\BIBdecl

\bibitem{bib:5G_arrays}
W.~{Hong}, Z.~H. {Jiang}, C.~{Yu}, J.~{Zhou}, P.~{Chen}, Z.~{Yu}, H.~{Zhang},
  B.~{Yang}, X.~{Pang}, M.~{Jiang}, Y.~{Cheng}, M.~K.~T. {Al-Nuaimi},
  Y.~{Zhang}, J.~{Chen}, and S.~{He}, ``Multibeam antenna technologies for 5{G}
  wireless communications,'' \emph{IEEE Transactions on Antennas and
  Propagation}, vol.~65, no.~12, pp. 6231--6249, 2017.

\bibitem{bib:automotive_radar_arrays}
B.~{Ku}, P.~{Schmalenberg}, O.~{Inac}, O.~D. {Gurbuz}, J.~S. {Lee},
  K.~{Shiozaki}, and G.~M. {Rebeiz}, ``A 77–81-{GH}z 16-element phased-array
  receiver with $\pm {\hbox{50}}^{\circ}$ beam scanning for advanced automotive
  radars,'' \emph{IEEE Transactions on Microwave Theory and Techniques},
  vol.~62, no.~11, pp. 2823--2832, 2014.

\bibitem{bib:chammas}
M.~El-Chammas and B.~Murmann, \emph{Time-Interleaved ADCs}.\hskip 1em plus
  0.5em minus 0.4em\relax New York, NY: Springer New York, 2012.

\bibitem{bib:razavi}
B.~{Razavi}, ``The current-steering {DAC} [a circuit for all seasons],''
  \emph{IEEE Solid-State Circuits Magazine}, vol.~10, no.~1, pp. 11--15, 2018.

\bibitem{bib:olieman}
E.~Olieman, \emph{Time-interleaved high-speed {D}/{A} converters}, 2016.

\bibitem{bib:beauchamp}
D.~{Beauchamp} and K.~M. {Chugg}, ``Machine learning based image calibration
  for a twofold time-interleaved high speed {DAC},'' in \emph{2019 IEEE 62nd
  International Midwest Symposium on Circuits and Systems (MWSCAS)}, 2019, pp.
  908--912.

\bibitem{bib:daigle}
C.~{Daigle}, A.~{Dastgheib}, and B.~{Murmann}, ``A 12-bit 800-{MS}/s
  switched-capacitor {DAC} with open-loop output driver and digital
  predistortion,'' in \emph{2010 IEEE Asian Solid-State Circuits Conference},
  2010, pp. 1--4.

\bibitem{bib:aleriza}
A.~Dastgheib, ``Calibration {ADC} and algorithm for adaptive predistortion of
  high-speed {DAC}s,'' Ph.D. dissertation, Stanford University, 2013.

\bibitem{bib:poly_osc_interp}
A.~Janczak, \emph{Identification of Nonlinear Systems Using Neural Networks and
  Polynomial Models: A Block-Oriented Approach (Lecture Notes in Control and
  Information Sciences)}.\hskip 1em plus 0.5em minus 0.4em\relax Berlin,
  Heidelberg: Springer-Verlag, 2004.

\bibitem{bib:dnc}
S.~Dey, S.~C. Kanala, K.~M. Chugg, and P.~A. Beerel, ``Deep-n-{C}heap: An
  automated search framework for low complexity deep learning,'' \emph{arXiv
  e-print arXiv:2004.00974}, 2020.

\bibitem{bib:relu}
A.~F. Agarap, ``Deep learning using rectified linear units (relu),''
  \emph{arXiv preprint arXiv:1803.08375}, 2018.

\bibitem{bib:adam}
D.~Kingma and J.~Ba, ``Adam: A method for stochastic optimization,''
  \emph{International Conference on Learning Representations}, 12 2014.

\bibitem{bib:timing_errors}
K.~L. {Chan}, J.~{Zhu}, and I.~{Galton}, ``Dynamic element matching to prevent
  nonlinear distortion from pulse-shape mismatches in high-resolution {DAC}s,''
  \emph{IEEE Journal of Solid-State Circuits}, vol.~43, no.~9, pp. 2067--2078,
  2008.

\end{thebibliography}

\end{document}